\documentclass[sigconf]{acmart}

\renewcommand\footnotetextcopyrightpermission[1]{} 

\renewcommand\footnotetextcopyrightpermission[1]{} 

\usepackage{xcolor}
\usepackage{enumitem}
\usepackage{amsmath}
\newcommand{\bowen}[1]{}

\newcommand{\toolname}[0]{\textsc{DevOps-LLM-Bot}}

\usepackage{url}

\AtBeginDocument{%
  \providecommand\BibTeX{{%
    \normalfont B\kern-0.5em{\scshape i\kern-0.25em b}\kern-0.8em\TeX}}}

\setcopyright{acmcopyright}
\copyrightyear{2024}
\acmYear{2024}
\acmDOI{XXXXXXX.XXXXXXX}

\begin{document}
\bibliographystyle{unsrt}
\title{Automated DevOps Pipeline Generation for Code Repositories using Large Language Models}

\author{Deep Mehta, Kartik Rawool, Subodh Gujar, Bowen Xu}
\affiliation{%
  \institution{North Carolina State University}
  \city{Raleigh}
  \state{North Carolina}
  \country{USA}
  \postcode{27606}
}
\email{{ dmmehta2, khrawool, sgujar, bxu22 } @ncsu.edu}



\renewcommand{\shortauthors}{Deep, Kartik, Subodh, Bowen}

\begin{abstract}
\end{abstract}

\maketitle

\bowen{Notes:
deadline: Fri 8 Dec 2023 (https://2024.msrconf.org/track/msr-2024-data-and-tool-showcase-track)
}

Automating software development processes through the orchestration of GitHub Action workflows has revolutionized the efficiency and agility of software delivery pipelines. This paper presents a detailed investigation into the use of Large Language Models (LLMs) – specifically, GPT 3.5 and GPT 4 – to generate and evaluate GitHub Action workflows for DevOps tasks. Our methodology involves data collection from public GitHub repositories, prompt engineering for LLM utilization, and evaluation metrics encompassing exact match scores, BLEU scores, and a novel DevOps Aware score. The research scrutinizes the proficiency of GPT 3.5 and GPT 4 in generating GitHub workflows, while assessing the influence of various prompt elements in constructing the most efficient pipeline. Results indicate substantial advancements in GPT 4, particularly in DevOps awareness and syntax correctness. The research introduces a GitHub App built on Probot, empowering users to automate workflow generation within GitHub ecosystem. This study contributes insights into the evolving landscape of AI-driven automation in DevOps practices.

\section{Introduction}
\bowen{motivation, why it's an important problem?}

\bowen{usually introduction is the most difficult section to write, write down the logic flow before write the detailed evidence}

Automation has become integral to modern software development, and GitHub Action workflows stand as a pivotal tool in orchestrating streamlined software delivery pipelines. This paper delves into an in-depth exploration of Large Language Models (LLMs), specifically GPT 3.5 and GPT 4, to generate and evaluate GitHub Action workflows tailored for DevOps tasks. Our research methodology involves the curation of pertinent data from public GitHub repositories and the engineering of prompts to guide LLMs in workflow generation. Evaluation metrics, including exact match scores, BLEU scores, and a newly introduced DevOps Aware score, assess the fidelity of generated workflows to DevOps best practices.


Moreover, a pivotal contribution of this research is the development of a dedicated GitHub App. This App capitalizes on the capabilities of LLMs, allowing seamless workflow generation for any GitHub repository and immediate commit into the repository itself. This innovative tool empowers developers by automating the generation and integration of workflows directly into their projects, streamlining their software development lifecycle.

The outcomes of this research present promising advancements, particularly in the enhanced capabilities of GPT 4, exhibiting heightened DevOps awareness and improved syntactic accuracy. The GitHub App's development further signifies the practical application of AI-driven models in the realm of DevOps practices, albeit with identified limitations.

\section{Related Work}

In the work titled "Let’s Supercharge the Workflows: An Empirical Study of GitHub Actions" \cite{chen2021let}, an in-depth analysis is conducted on the usage patterns of GitHub Actions. Findings reveal that during the GitHub Actions configuration, developers are required to set multiple components, each presenting numerous alternatives. This complexity poses a challenge for less experienced developers in configuring a suitable GitHub Action Workflow. Furthermore, the study highlights the dependence of the GitHub Action configuration process on developers' personal experience, suggesting a need for assistive tools to facilitate the configuration of GitHub Actions.

This paper \cite{pujar2023automated} addresses the gap of domain specific language models by focusing on Ansible-YAML, a widely employed markup language for IT Automation. The proposed solution, Ansible Wisdom, is a transformer-based model, uniquely trained on a dataset containing Ansible-YAML. To comprehensively assess its performance, the paper introduces two novel performance metrics tailored for YAML and Ansible, aimed at capturing the distinctive characteristics inherent to this specialized domain.

\section{Method and Implementation}

In the section, we details the core design (i.e., prompt engineering) in Section~\ref{sec:prompt} and implementation in Section~\ref{sec:tool} of our method \toolname{}.

\subsection{Prompt Engineering}\label{sec:prompt}

For generating a YAML file for a given repository, ideally, the more information about the repository in the prompt, the better performance is expected to be achieved. However, in practice, due to the token limit and also the imperfection of the LLM, we integrate our expert knowledge as the feature of the task into the prompt.

\begin{figure}[ht]
    \centering
    \includegraphics[width=0.75\linewidth]{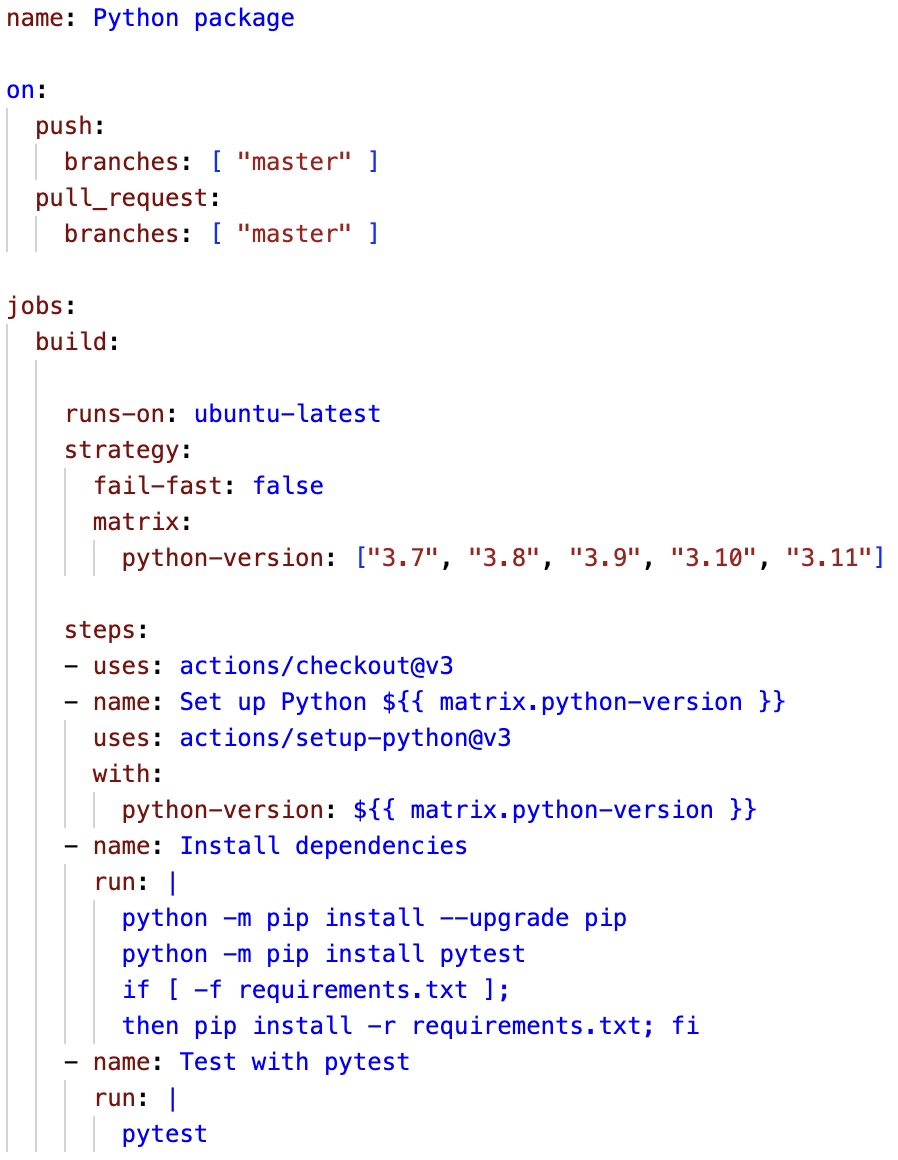}
    \caption{Example GitHub Actions Workflow}
    \label{fig:example-workflow}
\end{figure}

In this work, we customize the prompt with the following 2 parts as our first attempt. Particularly, for designing \emph{context}, we consider 2 pieces of task-specific information, which are \emph{file structure} and \emph{default branch} based on our observation. For example, Figure~\ref{fig:example-workflow} presents a real-world YAML file\footnote{\url{https://github.com/ajcr/rolling/blob/master/.github/workflows/python-package.yml}\bowen{add URL here}}. It carries both information about file structure (i.e. requirements.txt in the install step, \bowen{XXX in the Figure}) and branch (i.e. master while defining workflow triggers, \bowen{XXX in the Figure}).\bowen{pls provide a figure about a real world YAMl file that contains about file structure and branch info. It's very important to motivate/reason our prompt design.}

\subsection{Tool}\label{sec:tool}

\bowen{pls bold the text/description about all the customized part of the tool}

\begin{figure}[h]
    \centering
    \includegraphics[width=1\linewidth]{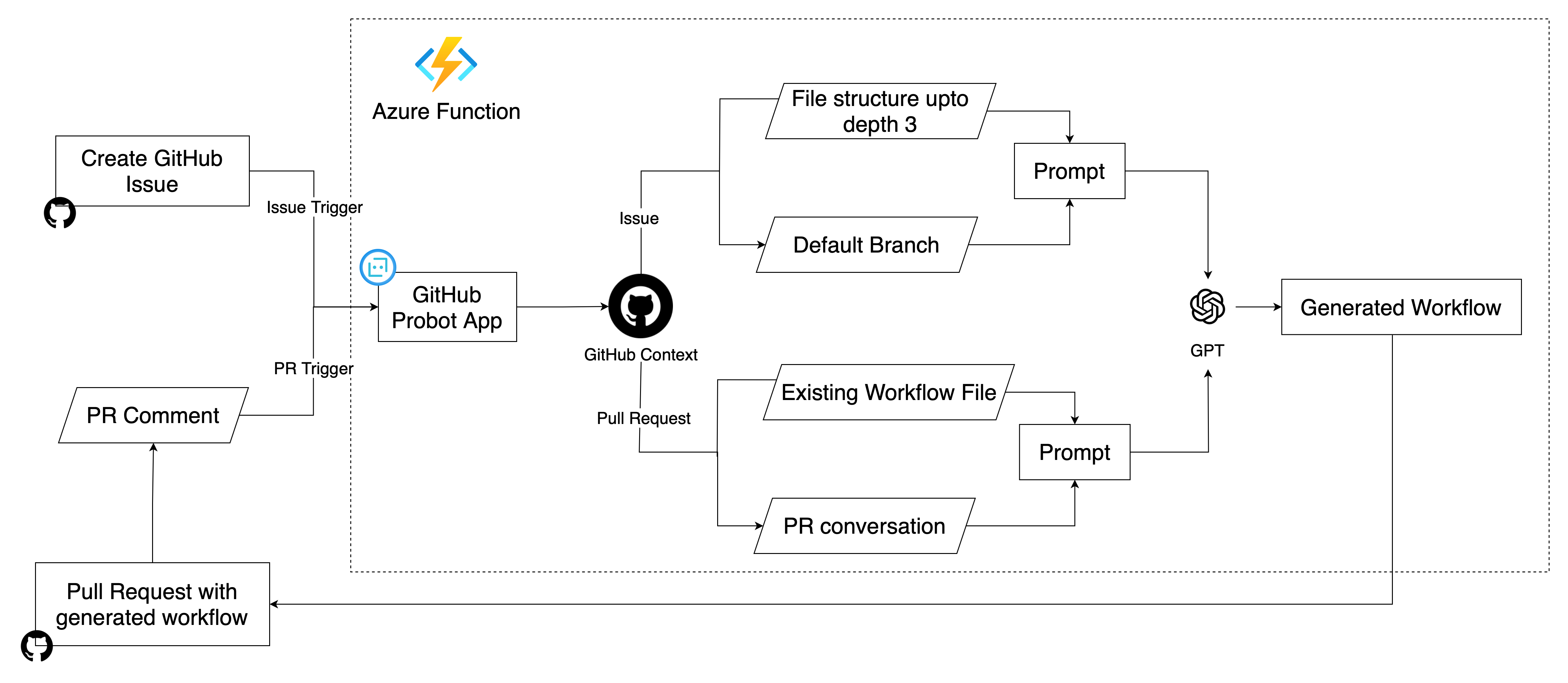}
    \caption{Architecture}
    \bowen{(1) the text in the fig is too small to be readable, pls increase the font size, (2) the description about this figure is completely missing in the main content.}
    \label{fig:architecture}
\end{figure}

We develop a tool based on GitHub App\footnote{\url{https://github.com/apps/devops-llm-bot}\bowen{pls add URL here}}. Figure \ref{fig:architecture} describes the architecture of this app built on top of Probot framework, which responds to the following events:

\begin{itemize}
    \item \textbf{Issue Creation}: The app springs into action when a new issue is created with a title that commences with \verb|@devops|. The context provided during this request includes the repository file structure and the default branch of the repository.
    \item \textbf{Pull Request Comment}: The app is also activated when a comment is made on a pull request, provided the comment starts with \verb|@devops-llm-bot|. In this scenario, the GPT is invoked with the context of the existing workflow and all previous conversations that the user has had within that pull request.
\end{itemize}

The app runs on \textbf{Azure functions} and uses the \textbf{Octokit} client to interact with GitHub and get information about the repository, such as its structure and contents. In our work, \textbf{we develop an app that invoked the GPT 4 (\emph{gpt-4-1106-preview}) chat completion API to generate workflow and raise a pull request for the developers to review, improvise by conversing with the bot and merge the workflow file}.


\bowen{minor: for each code block, can we add a line number at the beginning of each line?}


\bowen{the above exceeds the column width which is not allowed}

\subsection{Usage}

\begin{figure}[h]
    \centering
    \includegraphics[width=1\linewidth]{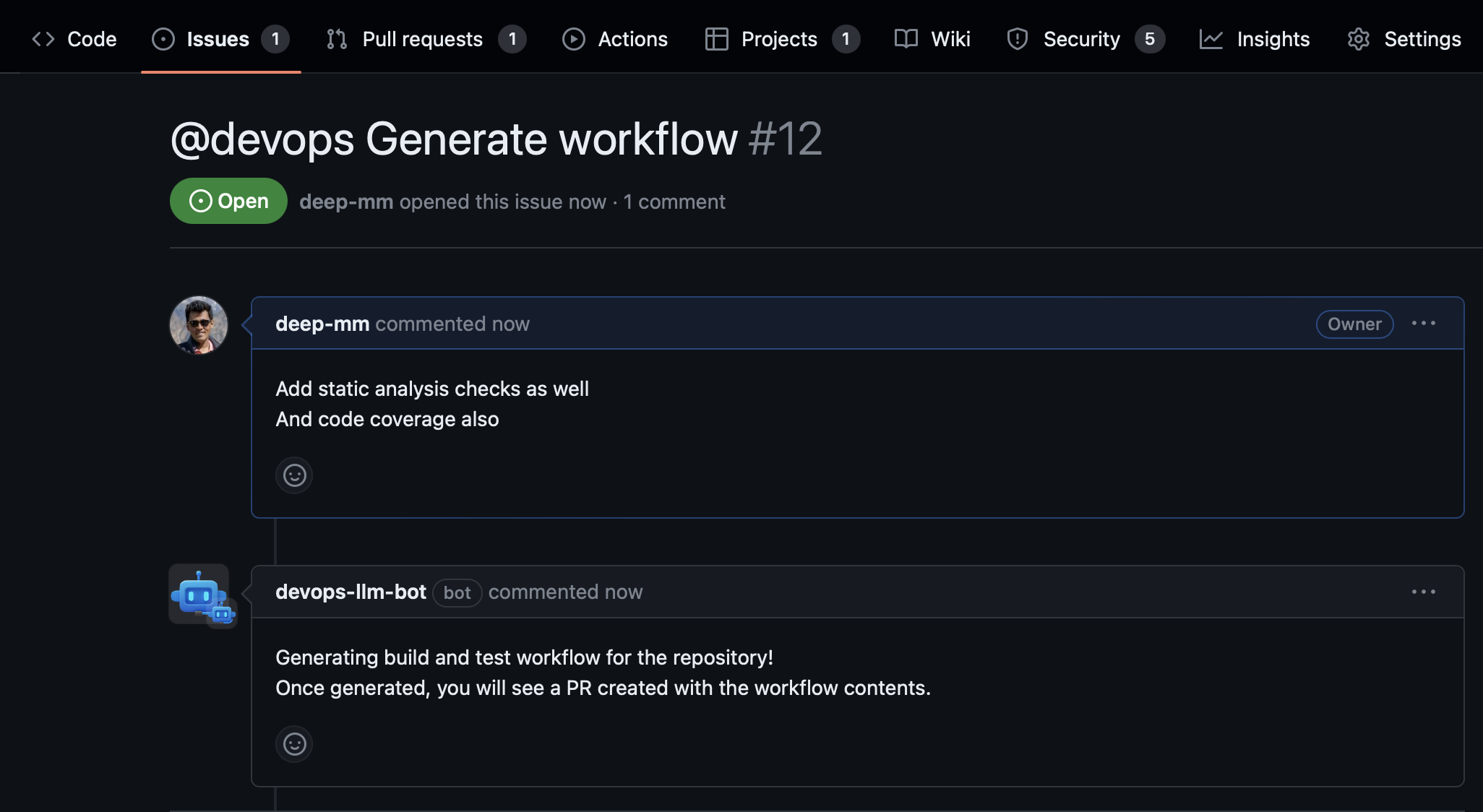}
    \caption{New GitHub Issue}
    \label{fig:output_1}
\end{figure}

As shown in Figure \ref{fig:output_1}, the bot can automatically generate a workflow for a GitHub repository when a user creates a new issue with a title that begins with \emph{@devops} and a body that specifies any custom requests. The bot replies to the issue with a message that indicates that it is working on the task. After the workflow is ready, the bot automatically commits it to a new branch and transforms the issue into a pull request that merges the new branch with the default branch of the repository.

\begin{figure}[h]
    \centering
    \includegraphics[width=1\linewidth]{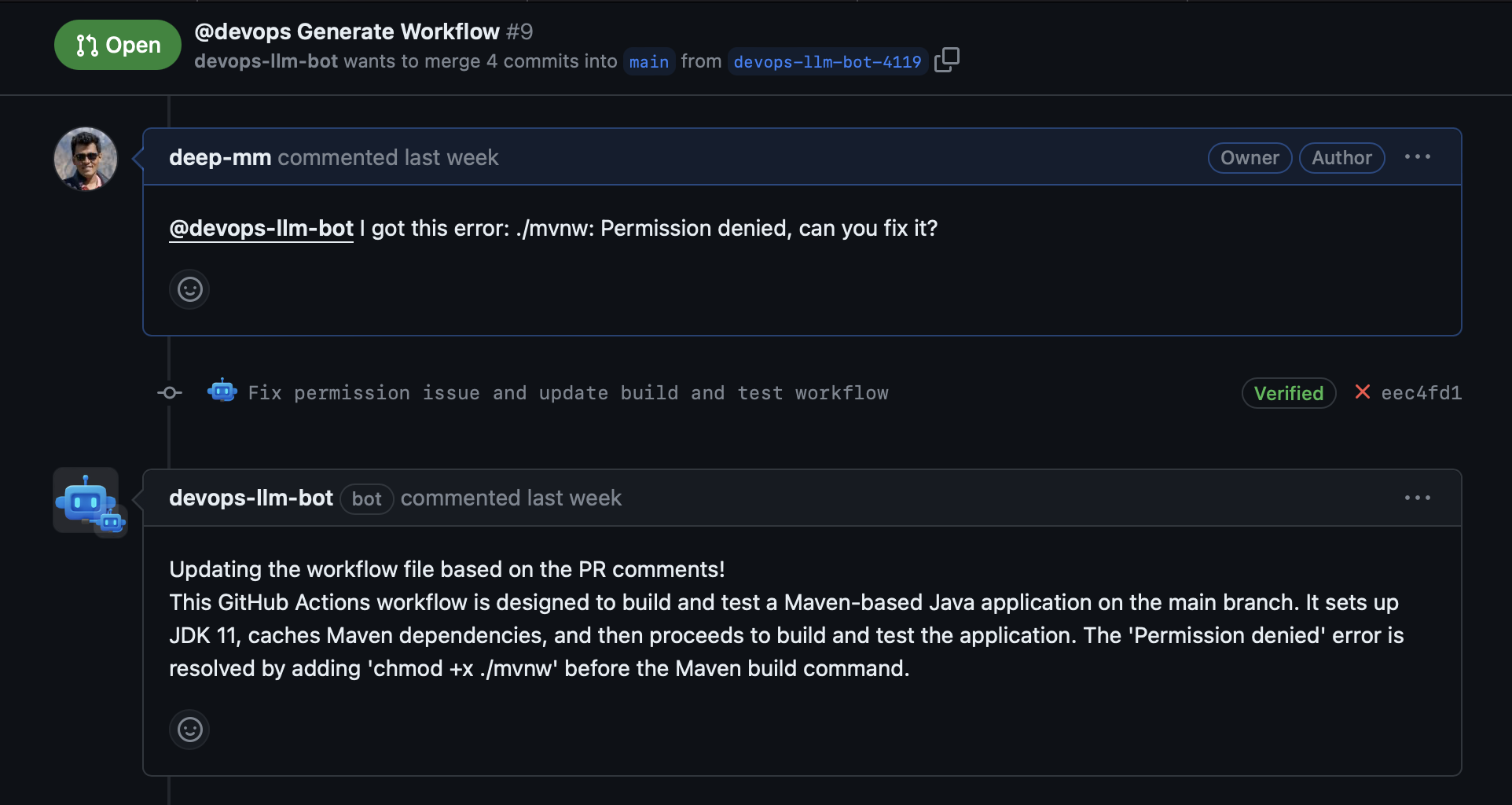}
    \caption{GitHub Pull Request Comment}
    \label{fig:output_2}
\end{figure}

Figure \ref{fig:output_2} demonstrates the direct interaction between the user and the LLM via the pull request, facilitating the necessary workflow updates. This interaction is designed to empower DevOps engineers by enabling them to generate the workflow without writing any code. It also allows them to verify the performance of the generated workflow, as it will automatically initiate upon the creation of the pull request. This ensures that the workflow operates as anticipated.



\section{Evaluation}
\subsection{Research Questions and Methodology}

Our experiment setting is designed to answer the following research questions.

\begin{itemize}
    \item How the performance \toolname{} differ based on different models?
    \item How do the different components of the prompt contribute to the final performance of \toolname{}?
    \item How efficient is \toolname{}?
\end{itemize}

\begin{figure}[h]
    \centering
    \includegraphics[width=1\linewidth]{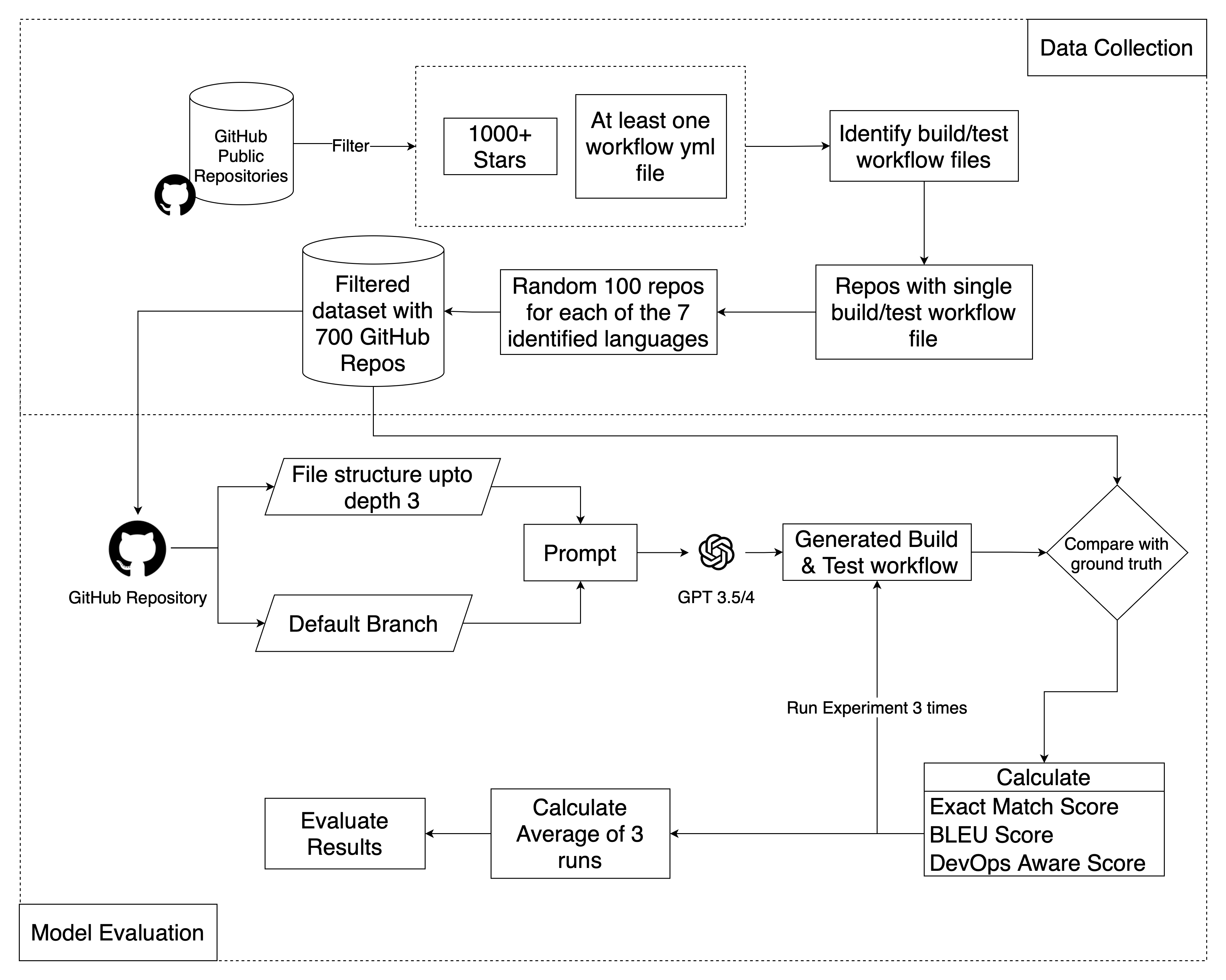}
    \caption{Framework for experimentation}
    \bowen{text font is too small, pls fix this}
    \label{fig:framework}
\end{figure}

Figure~\ref{fig:framework} presents the overview of our evaluation methodology. It demonstrates how we leverage public GitHub repositories to create a high-quality dataset for testing LLMs on DevOps tasks. In data collection (Section ~\ref{sec:data}), we apply filter conditions to select relevant repositories. In model evaluation (Section ~\ref{sec:metrics}), we generate build and test workflow with different settings in terms of prompt design and LLMs. We compare the generated workflows with the ground truth and evaluate them using multiple metrics in both automatic and manual manners.

\noindent\textbf{Automatic Evaluation}. 


We evaluate the pipeline based on the static similarity measurements between ground truth and the generated workflow. Specifically, we consider not only the exact match score, BLEU score, but also we introduce a customized metric DevOps Aware score (Section~\ref{sec:metrics}). To mitigate the randomness of the performance, we run each experiment three times for each setting and calculate the average scores.

Firstly, the syntax of the workflow is checked which checks if appropriate keys are used, the file follows workflow syntax and the structure is correct. Syntax validation of GitHub workflows is performed using actionlint \cite{actionlint} which is a static checker tool. 


For the evaluation, the workflow is checked for syntax if its valid or not and then the various metrics are calculated on the valid workflows.

\noindent\textbf{Human Evaluation:} In addition to automatic evaluation, we also conducted a manual evaluation to better understand and evaluate the quality of the generated workflow. Since we do not have a proven evaluation metric for yaml, a manual evaluation would help invalidate our novel DevOps Aware Score and also check how semantically our generated workflows are to ground truth. Hence, we sampled 324 generated workflows, with this sample size, we can maintain a confidence level of 95\% and a 5\% margin of error. The workflows were scored based on a 5-point Likert scale.

\subsection{Data Preparation}\label{sec:data}

To the best of our knowledge, our work is the first work that focuses on generating YAML file for a given code repository. Therefore, we construct the first dataset for the task with two stages, data collection and cleaning phase.

\noindent\textbf{Data collection phase}. We extract repository data from SEART-GHS tool (https://seart-ghs.si.usi.ch/) using specific criteria based on number of stars associated with each GitHub repository. The following filtering criteria is applied:

\begin{enumerate}
    \item \textbf{With at least 100 Stars}. To control the quality of our data, we apply the common indicator, the number of stars, to filter the low-quality repositories.
    \item \textbf{Contain at least 1 .yml or .yaml file within \emph{.github/} \emph{workflows} folder}. It indicates the presence of GitHub Actions.
\end{enumerate}



\noindent\textbf{Data cleaning phase}. By design, GitHub Actions can be implemented for various purposes, such as build, test, deployment, etc\footnote{\url{https://github.com/features/actions}}. Therefore, distinguishing the purpose of a specific GitHub Actions workflow file is challenging. To address this, we adopted an exhaustive list of build and test command lines tailored to specific programming languages. 



To streamline our research, we focused on a wide range of programming languages, they are Java, Python, JavaScript, TypeScript, Kotlin, C\#, \& C++. We select 100 data points for each language to construct a manageable dataset.



\subsection{Evaluation Metrics}\label{sec:metrics}

In our work, we not only consider the widely-used metrics such as Exact Match (EM), BLEU \cite{papineni2002bleu}, and Syntax Correct, but also introduce a new and specialized metric called \emph{DevOps Aware Score} to evaluate the generated workflow. The reason is that those traditional metrics carry certain known limitations. EM completely ignores the semantically equivalent yet different pairs between the generated one and the ground truth. BLEU score is another popular metric but it heavily relies on n-grams and may not capture the overall meaning of the YAML file. Syntax Correctness is also an important metric but it only focuses on the syntax-wise.

DevOps Aware is a customized metric defined specifically to compare the semantic distance between two YAML files. 
The metric aims to leverage knowledge of GitHub Action Workflow syntax to assess the effectiveness of code sections that directly impact the output. It acknowledges that achieving the same objective can be accomplished through various predefined actions. The evaluation process involves examining the ``jobs'' section, which comprises ``steps'' specifying actions and commands for execution. We extract the build/test ``jobs'' content from both the ground truth and the generated workflow, parsing through the "steps" to compare the generated steps with their corresponding actual steps. The other steps from the ground truth are overlooked because these workflows may encompass more than just building and testing. 

At present, our assessment is primarily concerned with the Language Learning Model's (LLM) ability to generate build and test actions. The comparison involves checking for keys such as ``uses,'' ``runs,'' or ``name.'' If both the generated step and the actual step contain "uses" or "runs," an Exact Match score is calculated. In other cases, the BLEU score is computed. The final score is the average of all the calculated scores. 


\begin{figure}
    \centering
    \includegraphics[width=0.75\linewidth]{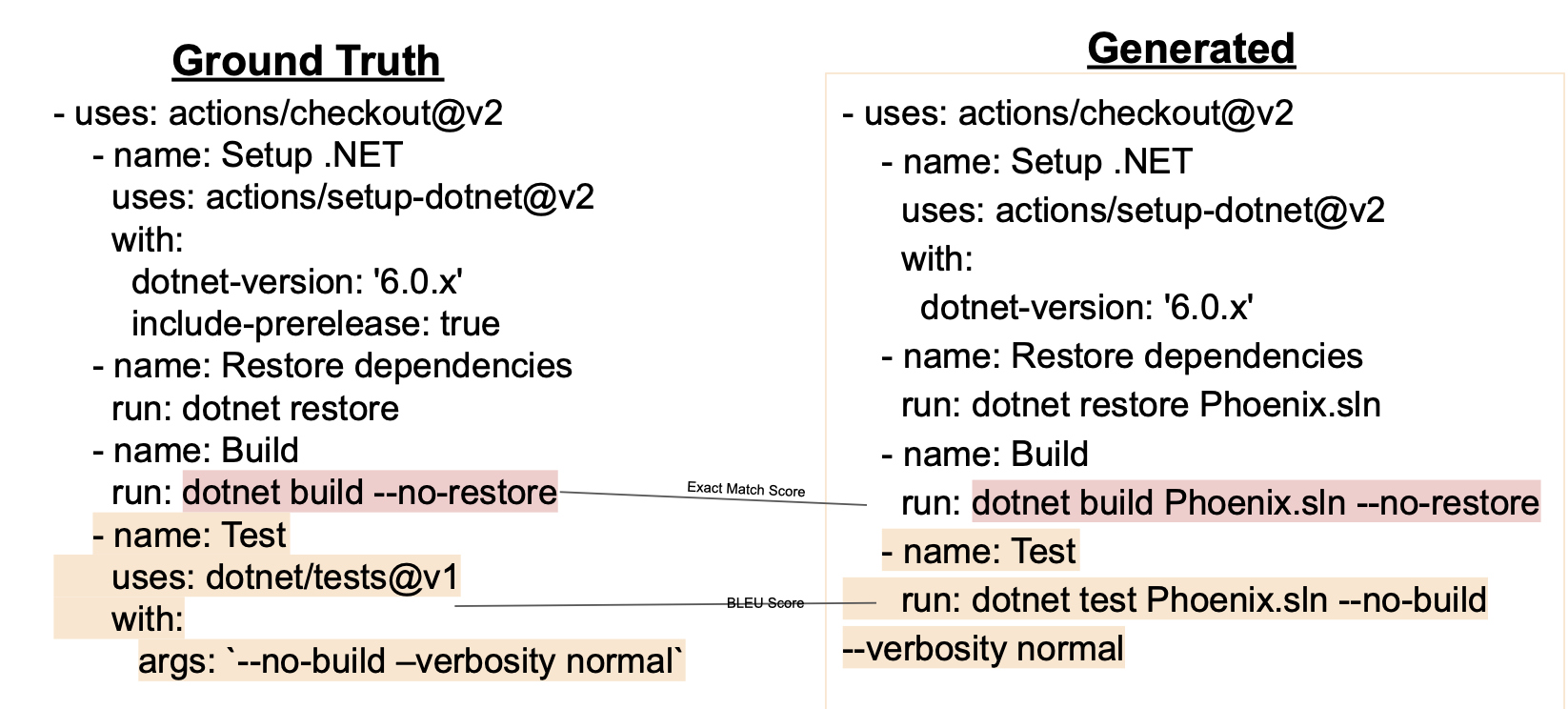}
    \caption{DevOps Aware Calculation}
    \bowen{every figure we draw in the paper, we should describe it.}
    \label{fig:devops-aware}
\end{figure}

\subsection{Result}

\textbf{Automatic Evaluation:}
In this comprehensive evaluation of GPT 3.5 and GPT 4 in generating GitHub Action workflows across multiple programming languages, several key metrics were employed to assess the models' performance. Notably, GPT 4 exhibited a marked improvement over its predecessor, GPT 3.5, in key areas such as DevOps awareness, BLEU score, and, syntax correctness. The DevOps Aware Score, a critical metric reflecting the models' understanding of DevOps practices in workflow generation, witnessed a substantial increase in GPT 4 across various languages, with notable improvements in C, C++, and Python. This suggests that GPT 4 has a more refined understanding of DevOps best practices, crucial for effective and seamless workflow automation.

\begin{table}[h]
\caption{Comparison of GPT 3.5 and GPT 4}
\label{tab:table1}
\begin{tabular}{ccccc}
 \toprule
Models  & DevOps  & BLEU & Exact  & Syntax Correct  \\
  & Aware Score & Score & Match & ( Valid / Total ) \\
\midrule
GPT 3.5 &           0.55         &        0.39    &      0.19       &    77.90 \%                \\
GPT 4   &           0.6      &      0.4      &        0.19     &      96.75 \%  \\
\bottomrule
\end{tabular}
\end{table}

\begin{table}[h]
\caption{Comparison of valid syntax percentage and 
DevOps Aware Score}
\label{tab:table2}
\begin{tabular}{ccccc}
 \toprule
Languages  & GPT 3.5  & GPT 4  & GPT 3.5  & GPT 4 \\
& Syntax & Syntax & Score & Score \\
\midrule
C\#&     87.33 \%    &    99.33 \%   &   0.44      &    0.53   \\
C++        &    72.00 \%     &     95.25 \% &     0.3    &    0.37  \\
Java       &     66.33 \%    &    97.6 \%  &     0.55    &    0.6  \\
Javascript &    88.00 \%     &    95.67 \%  &    0.7     &    0.74 \\
Kotlin     &     69.33 \%    &    97.32 \%  &    0.68     &    0.74  \\
Python     &    72.33 \%     &   95.00 \%  &    0.45     &    0.49    \\
Typescript &    90.00 \%     &    96.98 \% &    0.7     &   0.73  \\
\bottomrule
\end{tabular}
\end{table}

\newpage
\textbf{Manual Evaluation:} We have checked for the correlation between the manually evaluated workflows based on the 5-point likert scale and the DevOps Aware Score. We choose Pearson correlation for its capacity to quantify linear associations, enabling nuanced analysis of DevOps and manual score relationships. The overall correlation of 0.61 across all languages provides profound insights into the intricate dynamics between automated DevOps assessments and manual evaluations which validates our novel evaluation metric for DevOps workflows.




\bowen{describe the result from two aspect, automatic evaluation and human evaluation}

\section{How to use?}
We’ve developed a GitHub App as part of our research, which you can access [\href{https://github.com/apps/devops-llm-bot}{here}]. This app can be installed in any GitHub repository and used to generate GitHub build and test workflows for that repository.

You’re welcome to explore the bot with this [\href{https://github.com/deep-mm/myExpressApp/pull/68}{example PR}], or you can try it out in your own repository by following these steps:

\begin{enumerate}
    \item Install the GitHub app on your repository using the provided link.
    \item Create a new issue with a title beginning with @devops. The bot will use this to help generate your workflow. Any custom requests can be included in the issue description for the bot to consider while generating the workflow.
    \item You’ll see a comment on your issue from the bot, indicating that your request is being processed.
    \item After about a minute, the issue will transform into a pull request and your workflow file will be generated.
    \item If you’re satisfied with the workflow, you can merge the pull request.
    \item If you need modifications, add a comment in the pull request detailing the changes required. Make sure the comment begins with ‘@devops-llm-bot’ to trigger the bot to consider your changes and re-generate the workflow.
    \item Once the workflow is generated, the bot will reply on the PR and add a new commit. Enjoy exploring the bot! 
\end{enumerate}


\section{Conclusion}
In conclusion, the research outlined in this paper marks a significant milestone in the fusion of AI-driven capabilities, specifically GPT-4, within the realm of DevOps methodologies. The meticulous exploration and evaluation of GPT-4's capabilities showcased substantial advancements, particularly in its comprehension of DevOps practices and in generating syntactically accurate GitHub Action workflows.

The introduction of a dedicated GitHub App, empowered by Large Language Models, emerges as a significant contribution, allowing developers to automate workflow generation and seamlessly integrate it into their projects. This innovation streamlines the software development lifecycle, enabling developers to craft workflows efficiently and effectively.


Looking ahead, the future holds promising avenues for further enhancements. Fine-tuning models, experimenting with different models and prompt combinations, and extending the scope to include deployment steps within GitHub workflows are potential areas for future work. 



\newpage
\bibliography{ref}
\
\end{document}